\documentclass[prl,twocolumn, showpacs, showkeys,aps]{revtex4} 

\usepackage{amsmath}
\usepackage{amsfonts}
\usepackage{graphics}

\newcommand{\bra}{\left\langle}
\newcommand{\ket}{\right\rangle}

\newcommand{\psid}{\psi^{\dagger}}
\newcommand{\ret}{^{\text{(ret)}}}

\renewcommand{\Im}{\operatorname{Im}}

\begin{document}
\title{Luttinger liquid with asymmetric dispersion}
\author{Victoria I. Fern\'andez}
\author{An\'{\i}bal Iucci}
\author{Carlos M. Na\'on}
\email{victoria@fisica.unlp.edu.ar; iucci@fisica.unlp.edu.ar;
 naon@fisica.unlp.edu.ar}

\affiliation{Instituto de F\'\i sica La Plata. Departamento de
F\'\i sica, Facultad de Ciencias Exactas, Universidad Nacional de
La Plata. CC 67, 1900 La Plata, Argentina.\\ Consejo Nacional de
Investigaciones Cient\'\i ficas y T\'ecnicas, Argentina.}

\begin{abstract}
We present an extension of the Tomonaga-Luttinger model in which
left and right-moving particles have different Fermi velocities.
We derive expressions for one-particle Green's functions,
momentum-distributions, density of states, charge compressibility
and conductivity as functions of both the velocity difference
$\epsilon$ and the strength of the interaction $\beta$. This
allows us to identify a novel restricted region in the parameter
space in which the system keeps the main features of a Luttinger
liquid but with an unusual behavior of the density of states and
the static charge compressibility $\kappa$. In particular $\kappa$
diverges on the boundary of the restricted region, indicating the
occurrence of a phase transition.

\end{abstract}

\pacs{71.10.Pm}

\keywords{Luttinger liquid}

\maketitle

In the last years there has been much interest in the study of one-dimensional (1D)
condensed matter problems \cite{Reviews}. Specific examples of experimentally realized
1D structures are: strongly anisotropic organic conductors \cite{organic conductors},
charge transfer salts \cite{salts}, quantum wires \cite{quantum wires}, edge states in
a two-dimensional (2D) electron system in the fractional quantum Hall (FQH) regime
\cite{FQH} and the recently built Carbon Nanotubes \cite{CNT}. All these systems are
no longer described by the usual 3D-like Fermi liquid picture. They are believed to
belong to a novel, highly correlated state of matter known as the Luttinger liquid
(LL) \cite{LL}. Very recently, possible LL behavior in 2D high temperature
superconductors has also been reported \cite{Orgad}.

From the theoretical point of view the most widely studied 1D
model is the so-called ``g-ology" model \cite{Solyom}, which is
known to display the LL behavior characterized by spin-charge
separation and by non-universal (interaction dependent) power-law
correlation functions. In particular it predicts a momentum
distribution function that vanishes at $p_F$ as
$n(p)\sim(p-p_F)^{2\gamma}$, where $\gamma$ is related to the
strength of the electron-electron interaction (in the free case
one has $\gamma=0$ and $n(p)\sim\theta(p_F+p)$). One of the
simplest and yet very useful version of the ``g-ology" model is
the exactly solvable Tomonaga-Luttinger (TL) model, which
describes left and right-moving electrons subjected to
forward-scattering interactions \cite{TL}.

In this Letter we propose a simple modification of the TL model in which left and
right-moving electrons have {\em different} Fermi velocities $v_L$ and $v_R$. Previous
studies of LL systems involving more than one Fermi velocity are related to an special
class of chiral LL \cite{HC} and to multiband and multichain models \cite{multi}.
Another interesting problem in which one has different values for $v_{F}$ is the
interaction between parallel conductors leading to the so called Coulomb drag
\cite{drag}. We want to stress that the model we shall study is crucially different
from all these systems since it is neither a purely chiral LL nor a multiband system
with symmetric dispersion. Our theory is formally similar to a recently proposed model
for the study of spin-orbit coupling in interacting quasi-1D systems
\cite{spin-orbit}. These authors, however, concentrated their attention on the
interplay between velocity asymmetry and spin degrees of freedom, whereas here we
derive and analyze physical consequences connected to the asymmetric dispersion only.
As we shall see, this point of view allows us to obtain some novel non trivial
features of the system.

To be specific we start by considering an {\em asymmetric}
dispersion described by the following Hamiltonian

\begin{multline}\label{1}
H=-i\hbar\int
dx\,(v_R\psid_R\partial_x\psi_R-v_L\psid_L\partial_x\psi_L)\\ +
\pi U\int dx \,\psid_R\,\psi_R \,\psid_L\,\psi_L,
\end{multline}
where $\psi_{R,L}$ and $\psid_{R,L}$ are the electron operators
and $U$ is the strength of the forward-scattering
electron-electron interaction. In the ``g-ology" language we have
$g_2=\pi\, U$ and $g_4=0$. The extension of our results to the
general case ($g_4\neq0$) is straightforward, here we consider
this particular case in order to keep the discussion as clear as
possible. We will set $\hbar=1$ from now on. Please note that both
$v_L$ and $v_R$ are positive, and $U>0$ corresponds to repulsive
interactions. This is the case we shall examine throughout this
work.

Since the edge states of FQH systems have been successfully described in terms of
chiral fermions with drift velocities proportional to $E/B$ \cite{FQH} ($B$ is the
uniform transverse magnetic field and $E$ is an electric field that keeps electrons
inside the sample \cite{Fradkin}), the model above could be experimentally realized by
putting together the edges of two FQH samples in the presence of different fields such
that the resulting fractions are also different. In such experimental array $U$
represents the strength of the interaction between the charge-densities (CD) of each
fermionic branch. Recent experiments on tunneling between edge states of laterally
separated quantum Hall effect systems \cite{exp} seems to indicate that the experiment
we propose is indeed feasible.

One interesting result of this Letter is the appearance, due to the velocity
asymmetry, of a new available region in the space of couplings in which the model (1)
predicts an anisotropic phase, in the sense that the collective charge-density modes
associated to each branch propagate in the same direction. However one has to be
cautious with this prediction since our computations show that (1) is no longer valid
in this region. Indeed, as we shall see, if we define

\begin{equation}
v_0=(v_R+v_L)/2
\end{equation}
and

\begin{equation}
\epsilon=\frac{v_R-v_L}{v_R+v_L},\qquad \beta=U/2 v_0,
\end{equation}
we can only trust our model inside the unit circle in the
$\epsilon-\beta$ parameter space ($\epsilon^2+\beta^2=1$). We then
first explore the physics described by (1) in the restricted
region. We shall be specially interested in discussing the cases
of constant asymmetry ($\epsilon$ fixed) and constant interbranch
interaction ($\beta$ fixed). In so doing we found a drastic change
in the behavior of the charge compressibility $\kappa$ where the
value at zero asymmetry is multiplied by a factor which diverges
on the transition curve. Studied as a function of $\beta$ for
fixed $\epsilon$, it first reaches a minimum and then there is a
strong enhancement as $\beta\rightarrow\beta_0\equiv\sqrt{v_R
v_L}/v_0$, in opposition to the monotonous decay present in the
$\epsilon=0$ case. A similar change of behavior is present in the
density of states (DOS) function. For $\beta$ and $\epsilon$
sufficiently small one recovers an ordinary LL system
($v_0\rightarrow v_F$,$v_-\rightarrow -v_+$).

Outside the restricted region there is a change in the sign of one
of the ``plasmon" velocities, accompanied by a dramatic change in
the behavior of the Green function for that branch, which now
diverges at long distances. This, together with the fact that
$\kappa$ becomes negative, bring out that the model suffers some
kind of instability in this region. It is important to stress that
this region is absent for $v_R = v_L$.

We have studied the system (\ref{1}) by using functional
bosonization techniques \cite{FuncBos}. This amounts to defining
fermionic field operators in the Heisenberg picture. We then have
a field-theoretical, Lagrangian formulation of the model. This, in
turn, allowed us to obtain an action describing the dynamics of
the bosonic collective excitations of the system. Using this
action one can easily compute the dispersion relations for the CD
oscillations. For short-range, constant electron-electron
potentials, these dispersions are linear, with velocities given by
$v_\pm=v_0\eta_\pm$, and

\begin{equation}
\eta_{\pm}=\epsilon\pm\sqrt{1 - \beta^2}. \label{2}
\end{equation}
From this equation one sees that the propagation of the collective
modes takes place for $\beta<1$. It becomes apparent that, in
contrast to the usual answer for a TL model with $v_{R}=v_{L}$ and
$g_{4}=0$, here one has two different velocities $v_{+}$ and
$v_{-}$ for the propagation of left and right CD modes. Moreover,
one of the velocities $v_{+}$ or $v_{-}$ goes to zero as the
interaction and the asymmetry approach the curve
$\epsilon^2+\beta^2=1$ and changes its sign beyond that curve, as
anticipated above. If one keeps $\beta$ fixed this change of sign
occurs for $\epsilon=\epsilon_0\equiv\sqrt{1-\beta^2}$. At this
point, as we will see, there is a divergence in the charge
compressibility and in the DOS, which suggests that a phase
transition takes place.

Now, in order to get an insight into the physical consequences of the velocity
difference, we compute single-particle quantities: the Green function
$G_r(x,t)=\bra\psi_r(x,\tau)\psid_r(0,0)\ket_{\tau\rightarrow it}$ with
$r=+\,(R),-\,(L)$, the momentum distribution function, the spectral function
$\rho_r(q,\omega)$ given by

\begin{equation}\label{spectraldef}
\rho_r(q,\omega)=-\frac{1}{\pi}\Im G_r\ret(q,\omega),
\end{equation}
and the DOS defined as

\begin{equation}\label{densitydef}
N(\omega)=\frac{1}{2\pi}\sum_r\int dq\,\rho_r(q,\omega).
\end{equation}
In these equations, $G_r\ret(q,\omega)$ is the Fourier transform of the retarded Green
function:

\begin{equation}
G_r\ret(x,t)= i\theta(t)\bra\left\{\psi_r(x,t),\psid_r(0,0)\right\}\ket.
\end{equation}

For the normal phase ($\beta^2+\epsilon^2<1$), the Green function at $T=0$ is given by

\begin{multline}\label{3}
G_r(x,t)=\frac{1}{2\pi\alpha}
\left(\frac{\alpha}{\alpha-ir(x-v_rt)}\right)^{\gamma+1}\\ \times
\left(\frac{\alpha}{\alpha+ir(x-v_{-r}t)}\right)^\gamma
\end{multline}
where $\alpha$ is an ultraviolet cutoff. The constant $\gamma$ has the usual
expression in terms of the stiffness constant $K$, $\gamma=(K+K^{-1}-2)/4$, but in
this constant $v_F$ must be replaced by the average $v_0$:

\begin{equation}
K=\sqrt{\frac{v_0-U}{v_0+U}}=\sqrt{\frac{1-\beta}{1+\beta}}.
\end{equation}

In the outer region ($\beta^2+\epsilon^2>1$), for $v_{R}>v_{L}$, we get
\begin{equation}\label{4}
G_R(x,t)=\frac{1}{2\pi}\frac{\left[\alpha-i(x-v_-t)\right]^\gamma}
{\left[\alpha-i(x-v_+t)\right]^{\gamma+1}}
\end{equation}
and

\begin{equation}\label{5}
G_L(x,t)=\frac{-1}{2\pi\alpha^2}\frac{\left[\alpha-i(x-v_-t)\right]^{\gamma+1}}
{\left[\alpha-i(x-v_+t)\right]^\gamma}.
\end{equation}

From these results one obtains a momentum distribution of the Fermi type for the right
branch, i.e., $n_{R}(p)\sim\theta(p-p_R)$. However, the situation is very different
for $n_{L}$. Indeed, when taking the appropriate $t\rightarrow0^{-}$ limit in order to
employ the usual definition of $n_L(p)$, one finds that the correlator increases
linearly with distance, instead of having the $x^{-1}$ decay (typical of 3D-like
systems) as is the case for the right branch, or the $x^{-(2\gamma + 1)}$ behavior
that yields the LL result in the ``normal" LL region. This leads to
$n_{L}\sim\delta'(p+p_L)$. The appearance of this divergent (at long distances) left
correlator, together with a momentum distribution which is not a positive definite
quantity, are clear indications that the model given by (1) is unphysical beyond
$\beta^2+\epsilon^2=1$. (For $v_L>v_R$ the corresponding left and right behaviors are
exchanged). We then conclude that the model given by (1) yields sensible results for
$\epsilon^2+\beta^2<1$ and from now on we will restrict our study to that region.

From equation (\ref{spectraldef}) one can calculate the spectral
function $\rho_r(q,\omega)$. We obtain, as in the symmetric case,
only one singularity in the positive frequency sector and one in
the negative sector, as expected for spinless systems. The
function diverges at those points as

\begin{multline}
\rho_r(q,\omega)\sim (\omega-v_rq)^{\gamma-1}(\omega-v_{-r}q)^\gamma\\
\times[\theta(\omega-v_-q)\theta(\omega-v_+q)+\theta(v_-q-\omega)\theta(v_+q-\omega)].
\end{multline}
The exponents do not depend on $\epsilon$ whereas the position of the singularities
does.

Concerning the DOS we get

\begin{equation}
N(\omega)= N_0(\omega) \left(1-\frac{\epsilon^2}{\epsilon_0^2}\right)^{-1-\gamma}
\end{equation}
with

\begin{equation}
N_0(\omega)=\frac{1}{\pi v_0}
\frac{(\omega/\omega_0)^{2\gamma}}{\Gamma(2\gamma+1)(1-\beta^2)^{\gamma+1/2}},
\end{equation}
where $\omega_0=v_0/\alpha$ and $\Gamma$ is the Gamma function. We see that as the
asymmetry is increased, the DOS grows from its value at $\epsilon=0$, and diverges at
the point $\epsilon=\epsilon_0$. We want to stress that in systems with several
spectral branches as multicomponent Tomonaga-Luttinger model \cite{Reviews} and
spin-polarized Luttinger liquids \cite{kimura} a growing of the DOS as increasing the
velocity difference between spectral branches is also observed. This disagrees with
the result obtained in the system with spin-orbit coupling \cite{spin-orbit}.

Using standard linear response theory one can express the
conductivity as an integral of the retarded current density
correlation function. At this point one has to recall that the
naive definition of the current $j=v_R \rho_R - v_L \rho_L$ does
not satisfy the continuity equation \cite{Giamarchi}. This choice
for $j$ leads to a frequency-dependent conductivity that diverges
on the unit circle $\epsilon^2+\beta^2=1$ and becomes negative for
$\epsilon^2+\beta^2>1$. However, as shown in \cite{Giamarchi} it
is indeed possible to build a physical current starting from the
continuity equation. Extending this procedure for the present
$\epsilon\neq0$ case we obtain $j_{phys}=(1-\beta)(\rho_R -\rho_L)
+ v_0 \epsilon (\rho_R +\rho_L)$. Using this expression we were
able to get the frequency-dependent conductivity $\sigma$ as:
\begin{equation}\label{7}
\sigma(\omega) = \frac{v_0\,(1-\beta)}{\pi\,\omega},
\end{equation}
which is independent of $\epsilon$.

Let us now consider the static charge compressibility of this
system, defined as
$\kappa=\bra(\rho_R+\rho_L)(\rho_R+\rho_L)\ket(q,\omega)$ for
$\omega\rightarrow0$. A straightforward computation yields

\begin{equation}\label{8}
\kappa = \kappa_0\,
\left(1-\frac{\epsilon^2}{\epsilon_{0}^2}\right)^{-1}
\end{equation}
with
\begin{equation}
\kappa_0 = \frac{1}{\pi \,v_{0}\,(1+\beta)},
\end{equation}
where one sees the divergence that takes place, at fixed $\beta$,
for $\epsilon=\epsilon_0$. This is similar to the behavior of the
DOS, although both functions diverge with different exponents.
Note that beyond $\epsilon_0$ one obtains negative values for the
compressibility, a further indication that the model is not valid
in that region. In Figure 1 we show the dependence of $\kappa$ on
$\epsilon$ for different values of $\beta$. We see that the
asymmetry enhances the compressibility. Of course, it is also
possible to study $\kappa$ as function of the coupling $\beta$,
for a fixed asymmetry. This is depicted in Figure 2 where one sees
that in drastic departure from the symmetric case, which displays
decreasing $\kappa$ for increasing $\beta$, now $\kappa$ reaches a
minimum and then grows without bound as $\beta\rightarrow\beta_0$.

\begin{figure}
\includegraphics{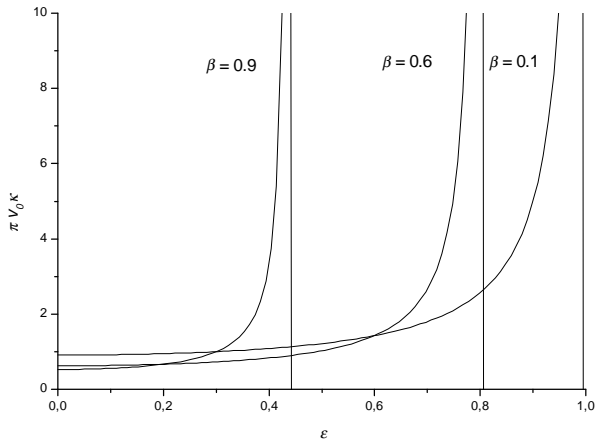}
\caption{\label{fig1} Static charge compressibility as function of
$\epsilon$, for $\beta=0.1, 0.6, 0.9$.}
\end{figure}

\begin{figure}
\includegraphics{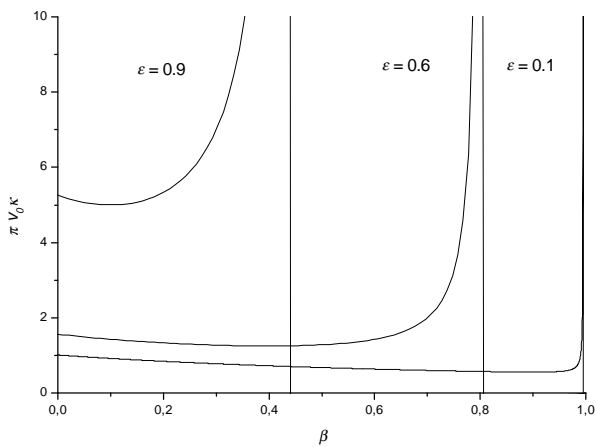}
\caption{\label{fig2}: Static charge compressibility as function
of $\beta$, for $\epsilon=0.1, 0.6, 0.9$.}
\end{figure}

The critical behaviour of the static charge compressibility and
the DOS at $\epsilon_0$ together with the ``freezing" of one of
the spectral branches ($v_{+/-}\rightarrow 0$) is an indication
that a phase transition involving CD degrees of freedom takes
place on the boundary $\beta^2+\epsilon^2=1$. A similar transition
related to spin variables was also found in \cite{spin-orbit}.

In summary, we have presented a simple modification of the usual
TL model, in which left and right-moving particles have different
Fermi velocities. By using functional bosonization methods we
computed the dispersion relations of the underlying bosonic
collective modes of the system. We showed that the velocity
asymmetry gives rise to some remarkable features. We have found
that the values for the electron-electron coupling $\beta$ and the
asymmetry $\epsilon$ are restricted to lie inside the
circumference $\epsilon^2+\beta^2=1$ in order to guarantee the
stability of the LL described by (1). In this region, the DOS
$N(\omega$) and the static charge compressibility $\kappa$ display
a big deviation from the standard LL behavior. For a fixed
asymmetry one sees that now $\kappa$ reaches a minimum and then
grows without bound as $\beta\rightarrow\beta_0$. On the boundary
of the restricted region the velocity of one spectral branch goes
to zero. Also $\kappa$ and DOS diverge when approaching this
critical boundary, indicating the appearance of a new phase.

To conclude we would like to stress that the idealized model we present here could be
approximately realized by allowing the interaction between the edge states of two FQH
plates. Since the drift velocities of the corresponding chiral LL's are proportional
to $E/B$, one could have $v_{R}\neq v_{L}$ by conveniently tuning up the corresponding
values of these fractions.

\begin{acknowledgments}
This work was partially supported by the Consejo Nacional de Investigaciones
Cient\'{\i}ficas y T\'ecnicas (CONICET) and Universidad Nacional de La Plata (UNLP),
Argentina. We are grateful to anonymous referees for valuable criticisms.
\end{acknowledgments}

\end{document}